\documentclass[aps,prc,twocolumn,times,graphicx,tighten,showpacs]{revtex4}
\usepackage[dvips]{graphicx}

\newcommand{\bea}{\begin{eqnarray}}
\newcommand{\eea}{\end{eqnarray}}
\newcommand{\be}{\begin{equation}}
\newcommand{\ee}{\end{equation}}
\newcommand{\ba}{\begin{eqnarray}}
\newcommand{\ea}{\end{eqnarray}}

\begin{document}
\title{Quasielastic Charged Current Neutrino-nucleus Scattering}

\author{J.E. Amaro}
\affiliation{Departamento de F\'{\i}sica At\'{o}mica, Molecular y Nuclear, 
Universidad de Granada,
  18071 Granada, SPAIN}
\author{M.B. Barbaro}
\affiliation{Dipartimento di Fisica Teorica, Universit\`a di Torino and
  INFN, Sezione di Torino, Via P. Giuria 1, 10125 Torino, ITALY}
\author{J.A. Caballero}
\affiliation{Departamento de F\'{\i}sica At\'{o}mica, Molecular y Nuclear,
Universidad de Sevilla,
  41080 Sevilla, SPAIN}
\author{T.W. Donnelly}
\affiliation{Center for Theoretical Physics, Laboratory for Nuclear
  Science and Department of Physics, Massachusetts Institute of Technology,
  Cambridge, MA 02139, USA}

\date{\today}

\begin{abstract}
We provide integrated cross sections for quasielastic
charged-current neutrino-nucleus scattering. Results evaluated
using the phenomenological scaling function extracted from the
analysis of experimental $(e,e')$ data are compared with those
obtained within the framework of the relativistic impulse
approximation. We show that very reasonable agreement is reached
when a description of final-state interactions based on the
relativistic mean field is included. This is consistent with
previous studies of differential cross sections which are in
accord with the universality property of the superscaling
function.
\end{abstract}






\maketitle

The development of present and future experimental studies of
neutrino oscillations at intermediate to high energies benefits
from high-quality predictions for neutrino-nucleus cross sections.
The kinematics involved in these processes typically lie in a
domain where a fully relativistic formalism is required; not only
should the reaction mechanism incorporate relativity, but also the
nuclear dynamics must be described in a relativistic framework.
Moreover, any reliable model of neutrino-nucleus scattering first
needs to be tested against electron scattering for similar
kinematics. Fully relativistic results for the latter have been
presented in the past~\cite{Udias,Jin,Chinn89} using the impulse
approximation, namely, where only one nucleon in the nucleus
interacts with the virtual photon exchanged in the process. Such
analyses have been shown to provide accurate descriptions of
quasielastic (QE) {\it exclusive} $(e,e'p)$ processes when a
proper description of the final-state interactions (FSI) between
the ejected nucleon and the residual nucleus is incorporated. This
is accomplished by using complex relativistic optical potentials
which have been fitted to elastic nucleon-nucleus scattering
data~\cite{Clark}. For {\it inclusive} processes of the type
$(e,e')$ and $(\nu,\mu)$, the contribution of all channels must be
retained; hence the use of complex potentials should be avoided
because of the loss of flux implied by the imaginary term.
Different approaches have been considered in the literature, e.g.,
the use of purely real potentials within the impulse
approximation~\cite{Chiara03,PRL,jac} and analyses based on the
Green function method~\cite{Horikawa1980,Giusti2003,Meucci:2006ir}.
Although both treatments lead to similar results, a comparison with
experimental QE cross sections is not yet conclusive because of
the effects introduced by ingredients beyond the impulse
approximation such as long- and short-range correlations,
meson-exchange currents (MEC) and, at high energies, the
excitation of the $\Delta$ resonance.

These difficulties can be partially overcome by making use of the
scaling behavior of the electron-nucleus cross sections through a
SuperScaling Analysis (SuSA) in the region of both the QE and
$\Delta$ peaks. Previous investigations of inclusive $(e,e')$
world data have clearly demonstrated the validity of scaling and
superscaling properties in these kinematical
domains~\cite{DS199,DS299,MDS02}. To summarize: by dividing the
experimental $(e,e')$ differential cross sections by an
appropriate single-nucleon factor one gets the superscaling
function, which embodies the basic information about the nuclear
dynamics. At sufficiently high energies this function depends upon
the transferred momentum ($q$) and energy ($\omega$) only through
a particular combination, the scaling variable
$\psi(q,\omega)$~\cite{DS299,MDS02} (first-kind scaling) and is
independent of the particular nucleus selected (second-kind
scaling).

Importantly, the superscaling function extracted from data
presents an asymmetric shape with a pronounced tail extending into
the region of high transferred energies, corresponding to positive
values of the scaling variable $\psi$. While this asymmetry is
largely absent in most non-relativistic models based on the
impulse approximation, the systematic study presented
in~\cite{PRL,jac} has shown that the correct amount of asymmetry
is provided in the relativistic impulse approximation (RIA) when
FSI are described with a relativistic mean field (RMF) potential.
Recently, a similar asymmetric scaling function has also been
obtained within a semi-relativistic (SR) approach including FSI
through a Dirac-equation-based model~\cite{semirel}.

The superscaling function $f(\psi)$ has been investigated in
detail using various models based on the impulse approximation.
The results are consistent with the existence of a {\it universal}
scaling function, which means that $f(\psi)$ is basically the same
for different types of reactions, such as $(e,e')$ and
$(\nu,\mu)$, provided that the same kinematical regime is
considered. As a consequence, instead of using specific models for
the nuclear structure and reaction mechanism, one can use the {\it
phenomenological} SuSA superscaling function extracted from
$(e,e')$ data to make reliable predictions for neutrino-nucleus
cross sections. This strategy was first pursued
in~\cite{neutrino1} for charge-changing (CC) $(\nu,\mu)$
processes, and has recently been applied~\cite{NC} to
neutral-current (NC) neutrino-nucleus scattering reactions. In
both cases, predictions were given for differential cross sections
and a comparison with various relativistic FSI predictions was
also provided for the two $t$-channel processes, $(e,e')$ and
$(\nu,\mu)$.

The differential cross sections and scaling functions were studied
at depth in~\cite{PRL,jac,neutrino1,neutrino2}, while in this work
we focus on integrated cross sections. We present and discuss
results for CC neutrino cross sections, integrated over the muon
energy as a function of the muon scattering angle, as well as for
cross sections integrated with respect to the scattering angle as
a function of the outgoing muon energy. Finally, the behavior of
the total cross section (integrated over both the scattering angle
and muon energy) as a function of the incident neutrino energy is
also investigated. We compare the results evaluated within the RIA
framework with those obtained by using the {\it phenomenological}
scaling function. We also include for reference the results
corresponding to the relativistic Fermi gas (RFG) model.

Studies of integrated cross sections are relevant for the analysis
of present and future neutrino oscillation experiments. The
significant differences observed in the differential cross
sections when comparing various model predictions~\cite{PRL,jac}
with the phenomenological scaling function can also have
consequences for the integrated cross sections. In particular, it
is of interest to determine how the large {\it asymmetry} in the
scaling function implied by the SuSA approach reflects on the
integrated cross section. A comparison with the different
theoretical descriptions, specifically the RMF approach which
reproduces the asymmetry of data, may give us important clues on
the validity of the diverse models.

In performing the analysis of integrated cross sections, an
important issue refers to some of the kinematical regions accessed
experimentally. In particular, small values of muon scattering
angle between the incoming neutrino and the outgoing muon,
$\theta_\mu$, imply small values of the momentum and energy
transferred in the process. In this situation, a theoretical
description of the reaction mechanism based on the impulse
approximation is questionable. Fortunately, the contribution at
these small angles to the global integrated cross section is
negligible because of the phase-space factor $\sin\theta_\mu$
entering the integral. 
Our aim here is to estimate the effects introduced in the
integrated cross sections by different theoretical models and to
compare these with those obtained by using the SuSA approach, and
thus we integrate over the full range of $\theta_\mu$ allowed by
the kinematics.

\begin{figure}[htb]
\begin{center}
\includegraphics[scale=0.55, bb=100 100 450 750]{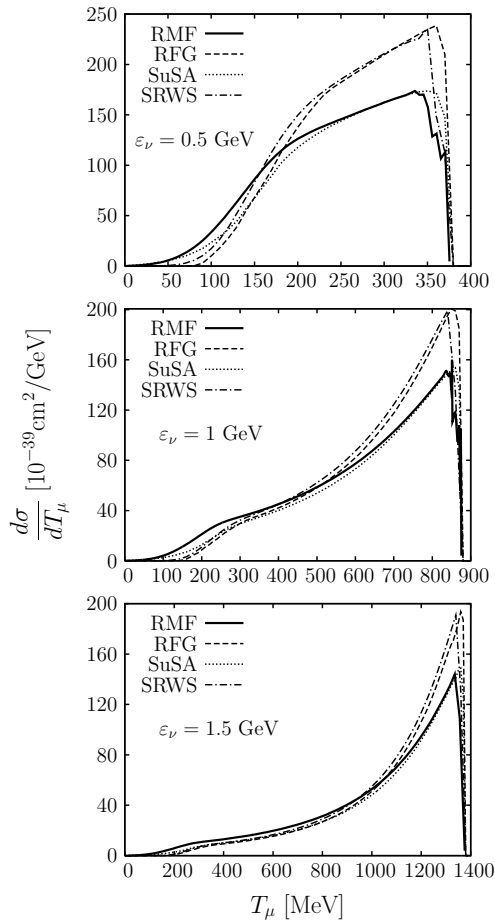}
\caption{CC cross section for $^{12}$C integrated over the muon
scattering angle as a function of the kinetic energy of the
emitted muon. Results for RMF (solid), RFG (dashed), SuSA
(dotted), and SRWS (dot-dashed) 
are compared for three different choices of kinematics.}
\label{CS_tmu}
\end{center}
\end{figure}


In Fig.~\ref{CS_tmu} we present results for the CC
$^{12}$C$(\nu_\mu,\mu^-)$ cross section integrated over the muon
scattering angle as a function of the muon kinetic energy.
Predictions corresponding to the relativistic impulse
approximation with FSI described with the RMF potential (solid
line) are compared with the cross sections evaluated by making use
of the phenomenological superscaling function $f(\psi)$ (dotted
line), labelled SuSA. We also show results obtained in the  RFG
model (dashed line) and in the SR approach described
in~\cite{neutrino2} using a Woods Saxon potential (dot-dashed
line) denoted as SRWS. The parameters of the WS potential have
been taken from \cite{neutrino2}, but with  the depth of the
central part of the proton potential adjusted to reproduce the
experimental mass difference between $^{12}$C and $^{12}$N. In the
RMF and SRWS approaches the experimental energies for the bound
neutrons are used, whereas in the SuSA and RFG approaches we have
taken into account the mass difference between the initial and
final nuclear systems, in accord with with Eq.~(6)
in~\cite{neutrino1}. This cuts out the higher muon kinetic
energies and produces a depletion of the cross sections.

Three different values of the incident neutrino energy have been
considered: $\varepsilon_\nu=$ 0.5, 1 and 1.5 GeV. The
relativistic plane-wave limit for the final nucleons, i.e., no FSI
considered (not shown in the figure), leads to cross sections
which are close to the RFG curves. Results in Fig.~\ref{CS_tmu}
show that the inclusion of FSI in SuSA and RMF gives rise to a
significant shift of strength: the cross section increases the RFG
result at lower muon energy values, whereas it quenches it at
higher energies. This reduction, similar for both RMF and SuSA
approaches, is about 20--25\% in the region close to the maximum.
In contrast, the SR model gives results that are similar to those
of the RFG. Note also that the use of real potentials for
describing the final nucleon states leads to the resonant
structure observed for high $T_\mu$ (that is, small energy
transfer $\omega$).

\begin{figure}[htb]
\begin{center}
\includegraphics[scale=0.55, bb=100 100 450 750]{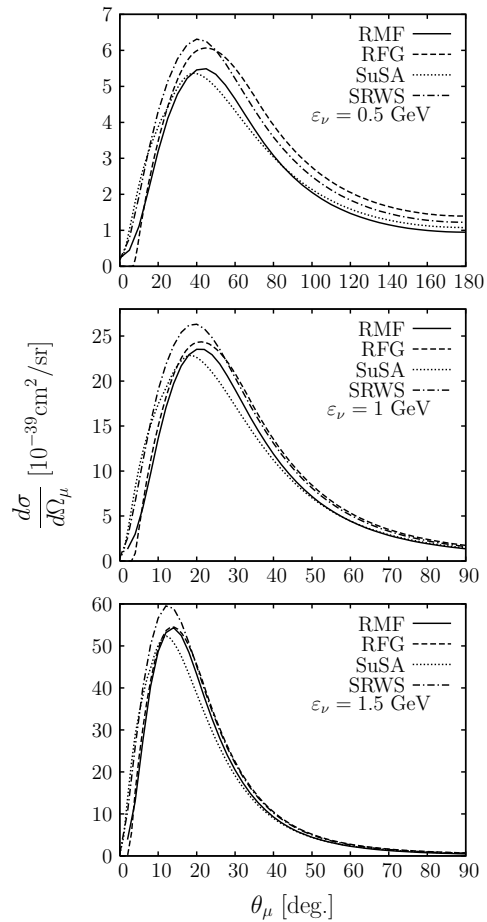}
\caption{Cross sections for the reaction $^{12}$C$(\nu_\mu,\mu^-)$
integrated over the emitted muon energy as a function of the
scattering angle $\theta_\mu$ for three values of the incident
neutrino energy $\varepsilon_\nu$. Results obtained within the context
of the RIA and FSI described through the RMF (solid lines) and SRWS
(dot-dashed) approaches are compared with the RFG model (dashed)
and with the cross section obtained by using the phenomenological
superscaling function extracted from the analysis of $(e,e')$ world
data, denoted by SuSA (dotted line).} \label{CS_thetamu}
\end{center}
\end{figure}


In Fig.~\ref{CS_thetamu} we present the cross section integrated
over the muon kinetic energy as a function of the muon scattering
angle. Again we compare results for RMF (solid line), RFG
(dashed), SuSA (dotted) and SRWS (dot-dashed). The inclusion of
FSI in RMF and SuSA gives rise to a depletion of the cross
section, except for very small angles.

As we show in the following, this implies a {\it smaller} global
integrated cross section, which means that, besides the
significant redistribution of strength produced by FSI, a global
reduction is also observed when an integral over the muon energy
and scattering angle is performed.

  From Figs.~\ref{CS_tmu} and \ref{CS_thetamu} it appears that the
SuSA and RMF approaches yield very similar results, corresponding
to a redistribution of strength with respect to the RFG and SR
models, in agreement with what was found in~\cite{PRL} and
\cite{neutrino2} for the double-differential cross sections. Note
also that nuclear model effects are more important for lower
neutrino energies ($\varepsilon_\nu=0.5$ GeV), in accord with
previous work~\cite{Chiara03,Alb97}, and that the scaling
arguments are only valid for high enough values of the transfer
momentum $q$, i.e., for high enough values of the neutrino energy.
Hence, some caution must be exercised when using the SuSA approach
for small $\varepsilon_\nu$.

\begin{figure}[htb]
\begin{center}
\includegraphics[scale=0.57, bb=100 450 450 750]{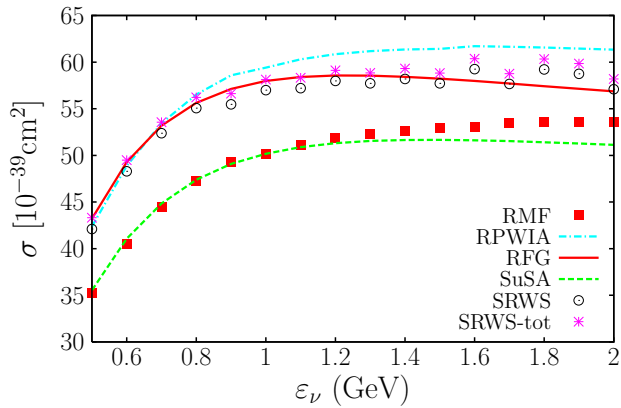}
\caption{(Color online)  
Total integrated CC cross section $\sigma$ for QE muon
neutrino reactions with $^{12}$C as a function of the incident
neutrino energy. We present results corresponding to RMF (squares), 
RFG (solid line), SuSA (dashed line),
RPWIA (dot-dashed line), SRWS (circles) and SRWS-tot (crosses).
}
\label{CS_total}
\end{center}
\end{figure}

Finally, nuclear model and FSI effects in the fully integrated
cross section are presented in Fig.~\ref{CS_total}, where the
total cross section $\sigma$ is plotted as a function of the
incident neutrino energy. Besides the four models considered in
the previous figures we also display the result corresponding to
RPWIA. All of the descriptions lead to a similar behavior for the
quasielastic cross section, which increases with the neutrino
energy up to $\varepsilon_\nu\sim$ 1--1.2 GeV and then saturates
to an almost constant value~\cite{Chiara03,Cris06}.

We observe that the RPWIA and RFG models give very similar results
up to $\varepsilon_\nu\simeq$ 0.8--1 GeV, while for higher
energies the RFG yields a lower cross section due to the
above-mentioned nuclear mass difference, which entails a cut in
the lower energy transfers. The SRWS model yields results which
are close to the RFG prediction. The results denoted SRWS-tot
include also the contribution of the discrete spectrum of $^{12}$N
obtained with the WS potential. This contribution turns out to be
very small (below 2\%). On the other hand, the RMF prediction
coincides with the SuSA one up to neutrino energies of about 1.2
GeV. Moreover, the FSI effects in RMF and SuSA remain sizeable
even at large $\varepsilon_\nu$. In fact, they tend to stabilize,
being of the order of $\sim$15\% at neutrino energies
$\varepsilon_\nu\geq 1.2$ GeV.

Before concluding, a comment is in order concerning the effect of
Pauli blocking (PB) in the RFG model. It is well-known that PB,
which is obviously accounted for in the RFG model, only affects
the low momentum and energy transfer region. However in the
process we are considering here this region is kinematically
forbidden due to the mass difference between the initial
($^{12}$C) and residual ($^{12}$N) 
nuclei. As a consequence the effect of PB in the integrated cross
sections turns out to be negligible.

Although no direct comparison of our predictions with experiment
can be performed, since no data on $^{12}$C are yet available in this energy 
range, the existing experimental data on $^2$H (which are, e.g., summarized 
in Ref.~\cite{Bodek}) are in qualitative agreement with the SuSA predictions. 
However, due to the above mentioned relevance of the mass difference bewteen 
the initial and final nucleus, a simple renormalization of the nucleon number
is not sufficient to test our model based on deuteron data.

In conclusion, we have evaluated the charged-current quasielastic
neutrino-nucleus cross sections integrated over the muon
scattering variables (kinetic energy and scattering angle) within
different relativistic theoretical approaches. Our results can be
summarized as follows:
(i) The effect of nuclear interactions is sizable for all values
of the neutrino energy ranging from 0.5 to 2 GeV and amounts to a
significant redistribution of the strength in the
single-differential cross sections and to a lowering by about
15--20\% of the total cross section with respect to the
relativistic Fermi gas result.
(ii) At high neutrino energies the differential cross section
$d\sigma/d\theta_\mu$ is strongly peaked at low scattering angles;
similarly $d\sigma/dT_\mu$ displays a pronounced maximum at high
muon kinetic energy. As a consequence the total cross section is
rather sensitive to the range of integration, which should be
carefully taken into account when comparing with experimental
data.
(iii) Finally, the integrated cross section evaluated with the
phenomenological superscaling function is very close to the RMF
prediction. This result complements our previous analyses of
differential cross sections and scaling functions, and it gives us
confidence in the adequacy of descriptions of QE $(\nu_\mu,\mu)$
reactions within the RIA context when FSI are included via strong
relativistic potentials.

\section*{Acknowledgements}
This work was partially supported by Ministerio de Educaci\'on y
Ciencia (Spain) and FEDER funds, under Contracts Nos. FPA2006-13807,
FPA2005-04460, FIS2005-01105, BFM2005-00810, by the Junta de
Andaluc\'{\i}a, and by the INFN-CICYT collaboration agreement
N$^o$ 05-22. It was also supported in part (TWD) by U.S.
Department of Energy under cooperative agreement No.
DE-FC02-94ER40818. The authors thank 
J.M. Ud\'{\i}as for providing the initial version of the RMF code and for
useful discussions.


\end{document}